\documentclass[%
 aip,
 amsmath,amssymb,
 reprint,%
]{revtex4-1}

\usepackage{graphicx}
\usepackage{dcolumn}
\usepackage{bm}
\usepackage{hyperref}
\usepackage[utf8]{inputenc}
\usepackage[T1]{fontenc}
\usepackage{mathptmx}

\begin{document}

\preprint{AIP/123-QED}

\title[Gating effects in antiferromagnetic CuMnAs]{Gating effects in antiferromagnetic CuMnAs}

\author{M. J. Grzybowski}
 \email{grzybowski@MagTop.ifpan.edu.pl}
 \affiliation{International Research Centre MagTop, Institute of Physics, Polish Academy of Sciences, Aleja Lotnikow 32/46, PL-02668 Warsaw, Poland}
 \affiliation{Institute of Physics, Polish Academy of Sciences, Aleja Lotnikow 32/46, PL-02668 Warsaw, Poland}
\author{P. Wadley}%
\author{K. W. Edmonds}
\author{R. P. Campion}
 \affiliation{School of Physics and Astronomy, University of Nottingham, Nottingham, United Kingdom}
\author{K. Dybko}
 \affiliation{International Research Centre MagTop, Institute of Physics, Polish Academy of Sciences, Aleja Lotnikow 32/46, PL-02668 Warsaw, Poland}
 \affiliation{Institute of Physics, Polish Academy of Sciences, Aleja Lotnikow 32/46, PL-02668 Warsaw, Poland}
\author{M. Majewicz}
 \affiliation{Institute of Physics, Polish Academy of Sciences, Aleja Lotnikow 32/46, PL-02668 Warsaw, Poland}
\author{B.~L.~Gallagher}
 \affiliation{School of Physics and Astronomy, University of Nottingham, Nottingham, United Kingdom}
\author{M.~Sawicki}
 \affiliation{Institute of Physics, Polish Academy of Sciences, Aleja Lotnikow 32/46, PL-02668 Warsaw, Poland}
 \author{T.~Dietl}
 \email{dietl@MagTop.ifpan.edu.pl}
 \affiliation{International Research Centre MagTop, Institute of Physics, Polish Academy of Sciences, Aleja Lotnikow 32/46, PL-02668 Warsaw, Poland}
  \affiliation{WPI-Advanced Institute for Materials Research, Tohoku University, Sendai 980-8577, Japan}


\date{\today}

\begin{abstract}
Antiferromagnets (AFs) attract much attention due to potential applications in spintronics. Both the electric current and the electric field are considered as tools suitable to control properties and the N\'eel vector direction of AFs. Among AFs, CuMnAs has been shown to exhibit specific properties that result in the existence of the current-induced spin-orbit torques commensurate with spin directions and topological Dirac quasiparticles. Here, we report on the observation of a reversible effect of an electric field on the resistivity of CuMnAs thin films, employing ionic liquid as a gate insulator. The data allow to determine the carrier type, concentration, and mobility independently of the Hall effect that may be affected by an anomalous component.
\end{abstract}

\maketitle

%




The interest in antiferromagnetic (AF) spintronics is stimulated by an increasing number of reports on different scenarios of manipulation of the N\'eel vector. Current-driven methods include spin-orbit torque commensurate with spin directions \cite{Zelezny:2014_PRL,Wadley:2016_Science,Grzybowski:2017_PRL,Wadley:2018_NN,Meinert:2018_PRA,Bodnar:2018_NC} or antidamping torque \cite{Chen:2018_PRL,Moriyama:2018_SR,Baldrati:2018_arXiv}. Although they provide means to control reversibly the N\'eel vector direction, they require a high current density. On the other hand, switching by an electric \emph{field} is considered promising for low-power spintronics. The electric field has been shown to modify a magnetic behavior of numerous ferromagnetic (FM) materials \cite{Ohno:2000_Nature,Boukari:2002_PRL,Chiba:2008_Nature,Sawicki:2010_NP,Chiba:2011_NM,Matsukura:2015_NN}, surprisingly including also rather conductive metal films \cite{Weisheit:2007_Science,Maruyama:2009_NN,Matsukura:2015_NN}, presumably because of an important role played by interfacial magnetic anisotropy \cite{Nozaki:2017_AM}.
It was also shown that an electric field can decrease the switching current in FM tunneling junctions \cite{Wang:2012_NM}. As for AFs, the electric field was proven to change a domain structure of multiferroic BiFeO$_3$ \cite{Zhao:2006_NM} or switch between AF and FM interactions in EuTiO$_3$ \cite{Ryan:2013_NC}. The magnetoelectric effect was also utilized to construct a memory device in $\alpha$-Cr$_2$O$_3$ film\cite{Kosub:2017_NC}. Metallic AFs were observed to exhibit modulation of the exchange spring effect \cite{Wang:2015_AM, Zhang:2016_SCPMA} or to change the magnetoresistance in AF-FM heterostructures \cite{Goto:2016_JJAP} due to the electric field. Finally, it has been recently demonstrated that it is possible to influence the spin-orbit torque by applying an electric field to a piezoelectric substrate \cite{Chen:2019_NM}. Furthermore, theoretical studies of CuMnAs show the coexistence of massless Dirac fermions and the AF order \cite{Smejkal:2017_PRL, Tang:2016_NP}. The reorientation of the N\'eel vector can induce the topological metal-insulator transition  \cite{Smejkal:2017_PRL, Tang:2016_NP}. However, to observe these phenomena experimentally, tuning the Fermi level to the band gap with an electric field would be highly desirable. In this context, exploring the influence of an electric field on CuMnAs is important from the point of view of low-power spintronics, topological aspects of AFs, and fundamental research on the role of electric fields in AFs.

In this paper, we demonstrate experimentally that the resistivity of highly conducting antiferromagnetic tetragonal CuMnAs thin films passivated by AlO$_x$ is reversibly modulated at room temperature by an electric field applied across an ionic liquid. The sign and the magnitude of the effect allow us to evaluate the carrier type, concentration, and mobility.  By comparing the field and Hall effect data we assess a possible magnitude of the anomalous Hall resistance. Conversely, under an assumption that the anomalous Hall resistance is negligible in collinear AFs, the consistency of the field and Hall effect results demonstrates that phenomena associated with surface charge trapping states \cite{Bauer:2012_NL}, electromigration \cite{Bauer:2015_NM}, and piezoelectricity \cite{Sztenkiel:2016_NC} are weak in CuMnAs, so that the main effect of gating is the formation of depletion and accumulation layers for  positive and negative gate voltages, respectively. The study yields also an upper limit for the dependence of the resistivity modulation on the direction of the current with respect to crystal axes.

\begin{table*}
\caption{\label{tab:hall}Properties of two CuMnAs films at room temperature determined from Hall measurements: the Hall coefficient $R_{\text{H}}$, hole concentration $p$  and mobility $\mu_{\text{H}}$ evaluated neglecting a possible contribution from the anomalous Hall effect.}
\begin{ruledtabular}
\begin{tabular}{ccccc}
CuMnAs layer thickness [nm] & $T$ [K] & $R_{\text{H}}\times10^8 [\Omega\text{cm}/\text{T}]$ & $p \times 10^{-21} [\text{cm}^{-3}]$  & $\mu_{\text{H}} [\text{cm}^2/\text{Vs}]$\\
\hline
45 & 300 & $4.1 \pm  0.1$ & $15.0 \pm 0.2 $ & $3.9 \pm 0.1$\\
10 & 283 & $13 \pm 4$ & $5 \pm1$ & $3.4\pm0.7$\\
\end{tabular}
\end{ruledtabular}
\end{table*}

\begin{figure}
\includegraphics{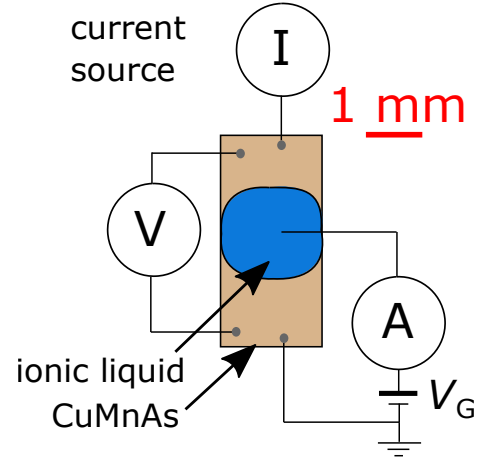}
\caption{\label{fig:1} Experimental setup for the determination of the resistivity changes under the influence of the gate voltage ($V_{\text{G}}$) for the device A. The blue circle denotes the area covered with the ionic liquid.}
\end{figure}
\begin{figure}
\includegraphics{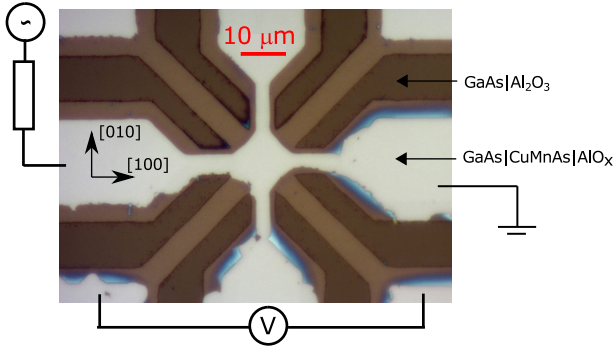}
\caption{\label{fig:2} Microdevice (device B) with eight contacts (clear bluish areas) for studies of the field effect for current along different crystalline axes. The darkest regions are trenches etched down to the substrate defining current paths, covered additionally by an Al$_2$O$_3$ film (brown areas) with atomic layer deposition to prevent chemical reactions between the layer and the ionic liquid. The Al$_2$O$_3$ film extends beyond the etched trenches covering partially the CuMnAs layer to ensure there is no contact with side walls and the ionic liquid. The whole device is covered by the ionic liquid in which the gate electrode is dipped. The current and voltage connections are shown for resistance measurements along the [100] crystal direction.}
\end{figure}

The field and Hall effect data have been obtained for a 10~nm CuMnAs tetragonal film grown coherently on a (001) GaAs substrate by molecular beam epitaxy, and capped with 2.5~nm Al layer that undergoes oxidation in the air. The thickness of the capping layer corresponds to the thickness of native Al oxide \cite{Evertsson:2015_ASS}. Additionally, Hall measurements have been carried out for a 45~nm film of CuMnAs grown on a (001) GaP substrate, and also capped with a 2.5~nm Al layer. Two devices have been prepared from the 10~nm film. The first one (device A) has been obtained from an elongated piece of the epilayer by fixing gold wires with silver paint and by depositing a droplet of the ionic liquid DEME-TFSI between them. As shown schematically in Fig.~\ref{fig:1}, about a half of the sample is cover by the gate. Another gold wire is dipped in the ionic liquid so that it does not touch the studied layer and forms a gate electrode. The microdevice B has been fabricated by means of multilevel electron beam lithography, wet etching, and lift-off to pattern four different current paths and eight gold contact pads, and by employing atomic layer deposition for growing an Al$_2$O$_3$ film serving to protect the etched trenches from undesired oxidation or chemical reaction with ionic liquid, as shown in Fig.~\ref{fig:2}. A wire bonder is used to fix wire probes to the contact pads. Similarly to the device A, the ionic liquid drop, deposited on the device top, and the gate electrode wire complete the field-effect structure. In the case of the microdevice B, the gate area is much larger than the central probed region.

The capacitance per area unit $C/S=(4.4 \pm 0.8)\times10^{-7}\,\text{F/cm}^{2}$ has been estimated for our ionic liquid by the $C$--$V$ profiling employing the frequency of 1\,kHz and the modulation voltage of 30\,mV superimposed on the d.c. voltage between 0 and 1\,V. This means that we can change interfacial charge density by about $3\times10^{12}$\,cm$^{-2}$ by the gate voltage of $V_{\text{G}}= 1$\,V.

The Hall resistance measured for our films is linear in the magnetic field in the studied range up to 9\,T and reveals a positive sign of the Hall coefficient, in agreement with earlier studies \cite{Wadley:2013_NC}.  Since magnetization of collinear AFs and, thus, spin polarization of band carriers vary also linearly with the magnetic field, the Hall resistance may {\em a priori} contain an anomalous component. Neglecting it, and adopting a single band transport model, our measurements lead to the values of hole concentrations $p$ and mobilities $\mu_{\text{H}}$, collected in Table~\ref{tab:hall}. The value $p = 1.1\times10^{22}$\,cm$^{-3}$ determined previously \cite{Wadley:2013_NC}, lies between $p = (1.50 \pm 0.02)\times10^{22}$ and $(0.5\pm0.1)\times10^{22}$\,cm$^{-3}$ obtained here for the 45 and 10\,nm thick films, respectively. A relatively small hole density in the thinnest layer, corresponding to the areal density of $5\times10^{15}$\,cm$^{-2}$, may point to interfacial or surface depletion. At the same time, the magnitudes of the Hall mobilities are within $3.6\pm0.3\,\text{cm}^2/\text{Vs}$ for the three films in question. A comparison of these values to the field mobilities will tell about the role of  surface states as well as to what extent the Hall data are affected by multiband transport and the anomalous Hall effect in the semimetallic and antiferromagnetic CuMnAs.

The key experiment of this work is the field effect, i.e., how the four-probe longitudinal resistance changes under the influence of the gate voltage. The main challenge is a high value of the carrier concentration in CuMnAs, making the magnitude of the field effect small and comparable to resistance changes caused by temperature fluctuations under ambient conditions. Therefore, a strong electric field has to be used and a good temperature stabilization implemented.  Accordingly, the studied devices are mounted on a sample holder in a vacuum chamber of a cryostat with a temperature controller, the arrangement preventing also a contact with water vapor in the air. The gate voltage is applied to the gate electrode in a form of a square wave with a period of 200 or 300\,s. In the case of the device A, the current source supplies a probing current $I$ in the range $1-100\,\mu\text{A}$ of alternating polarity (20~s period) to eliminate thermal forces.  In the case of the device B,  resistance changes generated by the gate voltage are probed by an ac lock-in method with an excitation current of $10\,\mu \text{A}$ and frequency $11\,\text{Hz}$. The device design allows to probe resistance along four different crystal axes, [100], [110], [010], and [1$\bar{1}$0].

\begin{figure}
\includegraphics{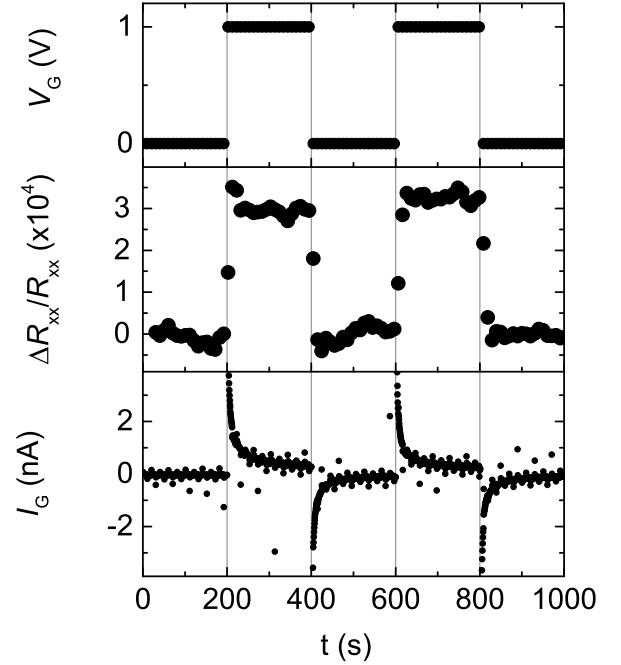}
\caption{\label{fig:3} Time dependence of the gate voltage $V_{\text{G}}$, relative resistance changes $\Delta R_{\text{xx}}/R_{\text{xx}}$ and current flowing through the gate $I_{\text{G}}$ for device A at room temperature for $10\,\mu\text{A}$ probing current in the experimental setup presented in Fig.~\ref{fig:1}. A clear correlation between changes of the gate voltage and longitudinal resistance can be seen, whereas the residual gate current $I_{\text{G}}$ shows  a different time dependence.}
\end{figure}

\begin{figure}
\includegraphics{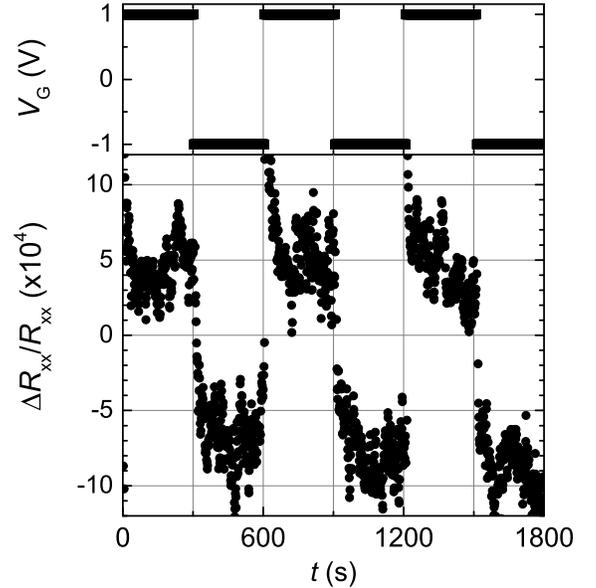}
\caption{\label{fig:devicebj100} Relative resistance changes $\Delta R_{\text{xx}}/R_{\text{xx}}$ for the device B under the influence of the gate voltage $V_{\text{G}}$ of the magnitude showed in the upper panel.  The current is flowing along the $[100]$ crystal direction.}
\end{figure}

As shown in Figs.~\ref{fig:3} and  \ref{fig:devicebj100}, clear variations of the resistance with the same periodicity as the gate voltage are observed for both devices. Assuming that neither electrochemical nor piezoelectric effects operate, an increase of resistance for positive values of the gate voltage means that hole carriers are involved.  The field-effect data are presented in the form of relative resistance changes $\Delta R_\text{xx}/R_\text{xx}$,
\begin{equation}
\frac{\Delta R_{\text{xx}}}{R_{\text{xx}}}=\frac{R_{\text{xx}}\left(V_{\text{G}}\right)-R_{\text{xx}}\left(V_{\text{G}}=0\,\text{V}\right)}{R_{\text{xx}}\left(V_{\text{G}}=0\,\text{V}\right)}, \label{eq:1}
\end{equation}
that is as a difference between the resistance when a specific gate voltage is on and when the gate voltage is zero, normalized by the resistance value at $V_{\text{G}}=0$. In the case of the device A, a small resistance drift linear in time is observed and subtracted from the data. It originates probably from a chemical reaction between the ionic liquid and edges of the sample. Its rate is $9 \times 10^{-5}\, \Omega/\text{s}$ for $V_{\text{G}}\!=\!1\,\text{V}$, but for most gate voltages used in the experiment it does not exceed $6 \times 10^{-5}\,\Omega/\text{s}$.

\begin{figure}
\includegraphics{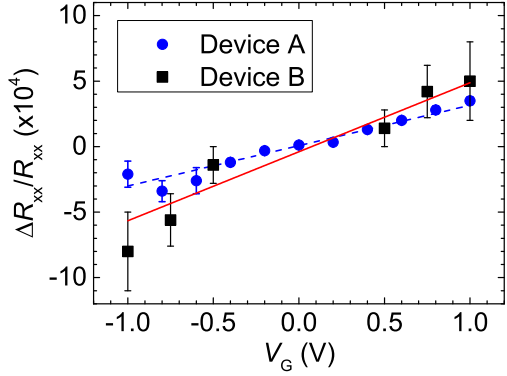}
\caption{\label{fig:5} Dependence of the relative resistance changes $\Delta R_{\text{xx}}/R_{\text{xx}}$ on the gate voltage $V_{\text{G}}$ for the two studied structures. The lines represent a linear fit to the experimental data for the device A (dashed, blue line) and device B (solid, red line).}
\end{figure}

\begin{figure}
\includegraphics{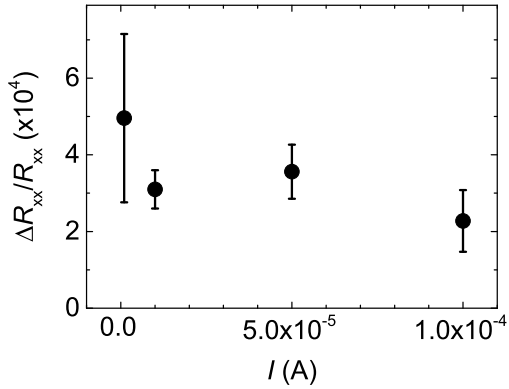}
\caption{\label{fig:6} Relative resistance changes $\Delta R_{\text{xx}}/R_{\text{xx}}$ of the device A recorded for different values of probing currents $I$ at $V_{\text{G}}\!=\!1\,$V. No apparent dependence is found in the studied current range, as expected.}
\end{figure}

The resistance changes depend on the magnitude of the gate voltage (Fig.~\ref{fig:5}), whereas they do not show any clear dependence on the probing current (Fig.~\ref{fig:6}). The current flowing through the gate $I_{\text{G}}$ (Fig.~\ref{fig:3}) decays with time. This dependence is presumably a long-time tail of two phenomena: (i) the capacitance charging effect via a non-zero resistance of the sample; (ii) a reorganization process of the charge distribution within the ionic liquid  \cite{Reichert:2018_FD,Jitvisate:2018_JPCL}. For the smaller device B the total current through the gate does not exceed $10\,\text{nA}$, which means that its magnitude in the probed region is much below 1\,nA.  The $V_{\text{G}}$ has been used in the range between -1$\,$V and 1$\,$V. A significant increase in $I_{\text{G}}$ and $R_{\text{xx}}$ for $V_{\text{G}}$ beyond this range suggests the onset of chemical reactions and electrical breakdown.

A linear fit to the experimental data for the device A presented in Fig.~\ref{fig:5} indicates that the relative resistivity change $\Delta R_{\text{xx}}/R_{\text{xx}}\!=\!(3.1\pm0.5)\times10^{-4}$ per 1\,V in the studied gate voltage range. The corresponding values for the device B are also shown for measurements carried out for different current directions with respect to crystal axes, and a linear fit to these data gives $\Delta R_{\text{xx}}/R_{\text{xx}}\!=\!(5\pm1)\times10^{-4}$ per 1\,V. Within the estimated experimental uncertainty there is no dependence of $\Delta R_{\text{xx}}/R_{\text{xx}}$ on the current direction. The lower value of $\Delta R_{\text{xx}}/R_{\text{xx}}$ in the case of the device A is assigned to only partial covering of the region between the voltage probes by the ionic liquid, as shown in Fig.~\ref{fig:1}. Actually, the data for these two samples are in accord if corrected by a factor $f$ describing a relative coverage of the probed area by the ionic liquid, where $f = 0.5\pm0.1$ and $f = 1$ for the device A and B, as shown in Fig.~\ref{fig:1} and Fig.~\ref{fig:2}, respectively.



We compare the experimental values of $\Delta R_{\text{xx}}/R_{\text{xx}}$ to theoretical estimations under the assumption that the only effect of the gate electric field is a depletion or accumulation of hole carriers at the layer surface. Under this assumption, a change in the areal hole density $\Delta p$ in the gated region is given by,
\begin{equation}
\Delta p= -\frac{CV_{\text{G}}}{Sq}, \label{eq:deltap}
\end{equation}
where $C/S = (4.4 \pm 0.8) \times 10^{-7}\,\text{F/cm}^{2}$ and $q =e$ for holes. On the other hand, assuming that the hole mobility $\mu$ is independent of the local carrier density as well as noting that in our case $\Delta p \ll pt$, where $t$ is the film thickness,
\begin{equation}
\frac{\Delta R_{\text{xx}}}{R_{\text{xx}}}= -f\frac{\Delta p}{pt}.
 \label{eq:deltar}
\end{equation}
For the areal hole concentration determined from the Hall measurements $(5 \pm 1) \times 10^{15}\,\text{cm}^{-2}$  we arrive to $\Delta R_{\text{xx}}/R_{\text{xx}}= f\cdot(5\pm1)\times10^{-4}$ per 1$\,$V, which is in good agreement with the experimentally observed values presented in Fig.~\ref{fig:5} for both devices taking into account the values of $f$ quoted above.

The Hall effect and the resistivity changes generated by gating allow us to compare the values of carrier mobility, namely the Hall mobility, $\mu_{\text{H}}\!=\!\sigma_{xx}/pq$, and the field mobility $\mu_{\text{E}}$ defined by
\begin{equation}
\mu_{\text{E}} = -\frac{1}{C/S}\frac{\partial \sigma_{\Box}}{\partial V_{\text{G}}},
\end{equation}
where $\sigma_{\Box}$ is the sheet conductivity in the gated region. In terms of the device longitudinal resistance $R_{xx}$, $\mu_{\text{E}}$ assumes the form,
\begin{equation}
\mu_{\text{E}} = \frac{L}{fWC/S}\frac{1}{R_{xx}^2}\frac{\partial R_{xx}}{\partial V_{\text{G}}},
\label{eq:mu}
\end{equation}
where $L$ and $W$ is the length and the width of the probed region, respectively; $L/W= 1.7\pm0.3$ and $1\pm0.1$ for the device A and B, respectively. The mobility values determined from the data in Fig.~\ref{fig:5} and Eq.\,\ref{eq:mu} for the studied structure are presented in Tab.~\ref{tab:mob}. 
The numbers quoted there imply that hole concentrations determined from the Hall resistance on the one hand, and from the field mobility and sample conductance on the other, are in accord.

\begin{table}
\caption{\label{tab:mob} Comparison of the Hall ($\mu_\text{H}$) and field mobility ($\mu_\text{E}$) for studied devices.}
\begin{ruledtabular}
\begin{tabular}{ccc}
 & $\mu_{\text{H}}\,[\text{cm}^2/\text{Vs}]$ & $\mu_{\text{E}}\,[\text{cm}^2/\text{Vs}]$\\
\hline
device A & - & $3.7 \pm 1$\\
\hline
device B & $3.4 \pm 0.7$ & $3.7 \pm 1$\\
\hline
\end{tabular}
\end{ruledtabular}
\end{table}

In summary, the electric field can modify reversibly the resistivity of CuMnAs structure capped with AlO$_{\text{x}}$. A quantitative  agreement between the values of the Hall and field mobilities proves that the modulation of the itinerant hole concentration in the layer is a mechanism accounting for the observed field effect. This points out that within the studied range of the electric fields electrochemical and piezoelectric phenomena as well as charging of surface states do not contribute significantly to the field-induced resistance changes, at least at room temperature. At the same time, there is no indications for a sizable contribution of the anomalous Hall effect. Similarly, there is no evidence for a breakdown of the single band approximation, as in the case of the multiband transport. In this situation the Hall effect would provide information of the highest mobility carriers, whereas the field effect on the band with the largest density of states at the Fermi energy.  The presented approach opens a way to manipulate the Fermi level and to explore the influence of the electric field on the magnetic order and the N\'eel vector switching in conducting antiferromagnetic systems. \\

The work was supported by the Polish National Science Centre (grants No. DEC-2016/21/N/ST3/03380 and DEC-2012/06/A/ST3/00247) by the Foundation for Polish Science through the IRA Programme financed by EU within SG OP Programme.
\nocite{*}
\bibliography{Electric_field_references}

\end{document}